# Influence of structural distortions on the Ir magnetism in $Ba_{2-x}Sr_xYIrO_6$ double perovskites

**Brendan F. Phelan\*, Elizabeth M. Seibel\*, Daniel Badoe Jr.\*, Weiwei Xie\*,**

**and R. J. Cava\***

\*Department of Chemistry, Princeton University, Princeton NJ 08540

Corresponding author: Brendan F. Phelan (bfp@princeton.edu)

**Abstract**

We explore the relative strengths of spin orbit coupling and crystal field splitting in the $Ir^{5+}$ compounds $Ba_{2-x}Sr_xYIrO_6$. In the case of strong spin orbit coupling and regular $Ir^{5+}$ octahedra, one expects a nonmagnetic $J = 0$ state; in the case of distorted octahedra where crystal field effects dominate, the $t_{2g}$ manifold splits into a magnetic ground state. We report the results of continuously transitioning from the cubic $Ba_2YIrO_6$ double perovskite with ideal octahedra to the monoclinic $Sr_2YIrO_6$ double perovskite with distorted octahedra. We see no emergence of an enhanced $Ir^{5+}$ magnetic moment in the series on increasing the structural distortions, as would have been the case for significant crystal field splitting. The near-constant magnetic moment observed through the $Ba_{2-x}Sr_xYIrO_6$ series reinforces the notion that spin-orbit coupling is the dominant force in determining the magnetism of iridium-oxygen octahedra in perovskite-like structures.

# 1. Introduction

Spin-orbit coupling (SOC) continues to be an important factor to consider when predicting and understanding the electronic and magnetic properties of strongly correlated materials that include heavy metals. It has proven to play an important role in the emergence of materials properties such as insulating antiferromagnets, superconductors, and topological insulators. [1] [2] [3] [4] [5] [6] [7] [8] [9] [10] [11] [12] Of particular interest have been extended systems of $IrO_6$ octahedra with iridium in a formal oxidation state of 4+. In these octahedral-$[Xe]5d^5$ systems, the five valence electrons occupy the triply degenerate $t_{2g}$ manifold. It is widely accepted that strong spin-orbit coupling will split this degenerate $t_{2g}$ manifold into a doublet $J_{eff}= 1/2$ and a quartet $J_{eff}= 3/2$ band. The completely filled, lower energy $J_{eff}= 3/2$ and the half-filled, higher energy $J_{eff}= ½$ bands give rise to an overall $J =1/2$ system. Much of the work studying these systems has focused on advanced spectroscopic techniques such as inelastic x-ray scattering. These studies point to the existence of a complex interplay between competing spin-orbit coupling and crystal field splitting, the latter of which is also able to break the degeneracy of the $t_{2g}$ manifold. [13] [14] [15] [16] [17] [18] [19] [20] With increasing distortion of the octahedra it is to be expected that the strength of crystal field will become a more important factor in determining the electronic structure of these compounds.

However, little work has focused on how far an octahedron can be distorted from the ideal shape before the electronic behavior and, as a result, how magnetism is affected due to the influence of the crystal field. For this purpose, systems featuring isolated $Ir^{5+}$ octahedra whose size or shape can be continuously varied are useful. Compared to systems featuring $Ir^{4+}O_6$ ($J_{eff}$=1/2) octahedra, those featuring $Ir^{5+}O_6$ ($J_{eff}$=0) octahedra remain scarcely studied through the lens of SOC. The expected spin state of these compounds should be zero in the point-charge picture and, as a result, small deviations of the crystal field resulting from structural distortions have the potential to leverage large changes in magnetic behavior.

For this reason we designed a study based on two previously reported compounds: $Sr_2YIrO_6$ [21] and $Ba_2YIrO_6$. [22] Ba, Sr, Y and O ions are all non-magnetic (they have closed electron shells) in this type of system. The former compound is monoclinic with significantly distorted and tilted $IrO_6$ octahedra, while the latter is a cubic compound with undistorted octahedra (Figure 1a and 1b, bond angles and lengths are taken from ref X). The structural characteristics of the $Ba_{2-x}Sr_xYIrO_6$ solid solution between these two end-member compounds have been reported in detail [23], and provide the crystallographic framework for the current work. By tracking the changes in the magnetic moment per Ir as the $IrO_6$ octahedra and the crystal structure become increasingly distorted, we can determine whether the structural distortions increase the crystal field enough to overcome the influence of the SOC and cause a significant change in the electronic structure.

## 2. Materials and methods

Polycrystalline samples of $Ba_{2-x}Sr_xYIrO_6$ ($0 \leq x \leq 2$, $\Delta x = 0.25$) were prepared by solid-state synthetic methods. Powdered samples were prepared from dried $BaCO_3$ (powder, 99.99%, Alfa Aesar), dried $SrCO_3$ (powder, 99.99%, Alfa Aesar), $Y_2O_3$ (nanopowder, 99.99%, Sigma-Aldrich) and Ir metal (powder, 99.95%, Alfa Aesar). Mixtures of starting materials were homogenized with mortar and pestle and then heated to 1200 °C under flowing $O_2$ for 36 hours. Samples were reground and heated again to

1300 °C under flowing $O_2$ for 36 hours. Samples were characterized using powder X-ray diffraction (PXRD) using a Bruker D8 ECO Advance with Cu Kα radiation and LYNXEYE-XE detector. [24]

Temperature dependent magnetizations (M) were measured with a Quantum Design Magnetic Property Measurement System (MPMS). Measurements of the M vs. applied magnetic field ($\mu_0H$) were linear up to applied fields of 1.5 T, and thus χ was defined as χ = $M/\mu_0H$, at an applied field of 1 T. Zero-field cooled (ZFC) measurements were performed on heating from 2 K to 300 K in a magnetic field of 1 T.

## 3. Calculations

Theoretical calculations on hypothetical model compounds in the $Ba_{2-x}Sr_xYIrO_6$ solid solution, using experimental $Ba_2YIrO_6$ structural data for $x$= 0-1.0, [22] and experimental $Sr_2YIrO_6$ structural data and for $x$= 1.25- 2. [21] Calculations were carried out using the Vienna Ab-initio Simulation Package (VASP) with the plane wave cutoff energy of 500 eV, and a set of 7×7×7 $k$ and 7×7×5 $k$ points for the irreducible Brillouin zones of the cubic $Ba_2YIrO_6$ –model and the monoclinic $Sr_2YIrO_6$–model, respectively. Exchange and correlation were treated by the generalized gradient approximation. To consider the effect of on-site repulsion for the 5$d$ element, Ir, LSDA calculations were performed with and without U, with $U$ = 4.7eV and $J$ = 0.7eV. The density of states (DOS) and band structure calculations for the compounds were performed by employing the Tight-binding Linear-Muffin-Tin-Orbital Atomic Sphere Approximation (TB-LMTO-ASA) using Stuttgart codes. Exchange and correlation were treated by the local density approximation (LDA) and the local spin-density approximation (LSDA). In the ASA method, space is filled with overlapping Wigner−Seitz (WS) spheres. The symmetry of the potential is considered spherical inside each WS sphere, and a combined correction is used to take into account the overlapping parts of the charge density. Empty spheres are necessary for this structure type, and the overlap of WS spheres is limited to no larger than 16%.

## 4. Results and Discussion

The solid solution of the type $Ba_{2-x}Sr_xYIrO_6$ ($0 \leq x \leq 2$, $\Delta x = 0.25$) was successfully prepared as single phase materials by the method described. [23] The X-ray powder diffraction patterns for the end members ($Ba_2YIrO_6$ and $Sr_2YIrO_6$) and one intermediate composition $BaSrYIrO_6$ are shown as examples in Figure 1c. In order to experimentally determine the effective moment ($\mu_{eff}$) per iridium for the members of the solid solution, we measured the zero field cooled (ZFC) magnetization versus temperature at an applied field of 1 T from 2 to 300 K. The temperature dependent magnetic susceptibilities ($\chi$) are shown in Figure 2a, b, and c, plotted in different ways. Figure 2a shows the general behavior of the susceptibility on a linear scale; it can immediately be seen that there is no dramatic increase in $\chi$ as the $IrO_6$ octahedra and the inter-octahedra bond angles become more distorted with increasing x. Thus the raw data show that the system does not become substantially more magnetic with increasing structural distortion. The inset to Figure 2a shows the measured susceptibilities at high temperatures for all samples studied. From these data we later extract the susceptibility at 250 K as a measure of the overall paramagnetism of each material. At high T this is dominated by a temperature independent term ($\chi_0$) that is extracted from the fits, as described.

The effective magnetic moments ($\mu_{eff}$) per iridium were calculated from the Curie constant ($C$, $\mu_{eff}=\sqrt{8C}$) extracted by fitting plots of inverse susceptibility ($1/(\chi-\chi_0)$) versus T (Figure 2b) to the Curie Weiss law $\chi-\chi_0 = C/(T-\theta)$, where $\theta$ is the Weiss temperature and T is temperature in Kelvin. This figure shows that there are two regions in the $1/(\chi-\chi_0)$ plot. At low temperatures, below 20 K, the straight lines in the plot (expanded in Figure 2c) show intercepts that are near 0 K. This is the intrinsic behavior of the magnetic system in its ground state. At higher temperatures, the curves turn over, and somewhat higher moments and higher apparent Curie Weiss temperatures can be inferred for the whole series. However, this high temperature regime is not appropriate for detailed analysis, as it is influenced by excitations out of the magnetic ground state similar to those frequently seen in rare earth systems, leading to a misunderstanding of the behavior if not considered. [25] [26] Figure 2c shows the low temperature regime, below 20 K, that is fit to determine the magnetic moment per iridium, and the Weiss thetas of the system.

The magnetic data for the $Ba_{2-x}Sr_xYIrO_6$ series are summarized in Figure 3. The $\mu_{eff}$, $\theta$, and $\chi_0$ values obtained from the fits as well as the raw $\chi_{250}$ data are shown. The figure quantitatively shows what is apparent from inspection of the raw data: the magnetic characteristics of the system do not change significantly across the series. For example there is remarkably little variation of the measured magnetic moments, which are within 0.1 of the average value of $\mu_{eff} = 0.47$ for the whole series. The observed susceptibilities, which are much lower than those observed by Cao et al, [27] are consistent with those observed by Kennedy et al, [23] (although our interpretation of the magnetic data is different, a more conventional, Curie-Weiss interpretation, than appears in that latter work). We observe that the Weiss thetas are very small in all cases, varying between about -5 and 0 K, as expected for materials in which $IrO_6$ spin centers are surrounded by non-magnetic octahedra (and thus magnetically isolated) and where small moments are present. While it is expected from a simple point charge model that strong SOC interactions will cause $Ir^{5+}$ to have all paired spins ($J_{eff}=0$) and therefore a net $\mu_{eff}=0$, it is not uncommon for $Ir^{5+}$ to exhibit non-zero moments experimentally. [28] Finally, it can also be seen from the figure that the high temperature susceptibility, characterized by $\chi_{250}$ and mirrored in the $\chi_0$ values obtained from the fits, shows relatively little variation across the series. There is a small broad peak in both quantities near x = 1.25, but whether this is meaningful is not clear at the current time.

To gain a fuller picture of how the experimental results agree with theory, the calculated moments are also plotted Figure 2c. When compared to the experimental moments we find that the theoretical calculations support our experimental data. Thus the conclusion is that distortions of the shape and bond angles of the $IrO_6$ octahedra in the $Ba_{2-x}Sr_xYIrO_6$ system do not generate crystal field splitting strong enough to quench the competing SOC. It is of interest that the simple electronic structure calculations employed here, in which only a basic electronic model have been applied, yield moments that are consistent with the experimental observations. We expect that recent, much more sophisticated calculations [29] will yield more detailed insight into the electronic and magnetic behavior of this system.

## 5. Conclusion

Investigating the interplay between crystal field effects and SOC has proven to be an area of key importance in the overall study of Iridium based oxides. The system we have presented here, with its isolated IrO$_6$ octahedra and familiar double perovskite structure, further confirms the dominance of spin orbit coupling in determining the electronic and magnetic properties of these materials.

## Acknowledgements

This research was supported by ARO MURI grant number W911NF-12-1-0461.

**Figure Captions:**

**Figure 1: a)** The crystal structures of $Ba_2YIrO_6$ and $Sr_2YIrO_6$. The $IrO_6$ octahedra in $Ba_2YIrO_6$ are regular, with 6 Ir-O bond lengths of 1.99 Å and O-Ir-O angles of 90°. [22] The $IrO_6$ octahedra of $Sr_2YIrO_6$ are significantly distorted, with Ir-O bond lengths between 1.95-2.00 Å, and O-Ir-O angles of 85.4°-94.6°. [21] **b)** Representative powder X-ray diffraction patterns of $Ba_2YIrO_6$, $BaSrYIrO_6$ and $Sr_2YIrO_6$.

**Figure 2:** Plot of the observed magnetic susceptibility ($\chi$) vs Temperature (T) for $Ba_{2-x}Sr_xYIrO_6$. Insert: detail of the higher temperature region.

**Figure 3: a)** Plot of observed inverse susceptibility ($1/(\chi - \chi_0)$) vs Temperature (T) for $Ba_{2-x}Sr_xYIrO_6$. **b)** Expansion of the low temperature region of ($1/(\chi - \chi_0)$) vs Temperature (T) for $Ba_{2-x}Sr_xYIrO_6$. Lines are from the fits, indicating Curie-Weiss behavior.

**Figure 4:** Plot of Weiss temperature $\theta$ (empty squares), $\chi$ at 250 K (empty triangles), $\chi_0$ used for fitting (filled triangles), measured moment $\mu$ (filled squares), and calculated moment $\mu$ (black x). The lines are provided as a guide to the eye.

**Figure 5:** Electronic structure calculation for $Ba_2YIrO_6$ without spin-polarization **a)** Density of States and **b)** Band structure, green indicates the partial DOS of Iridium.

**Figure 6:** Electronic structure calculation for $Sr_2YIrO_6$ without spin-polarization **a)** Density of States and **b)** Band structure, green indicates the partial DOS of Iridium.

**Figure 1:**

a)

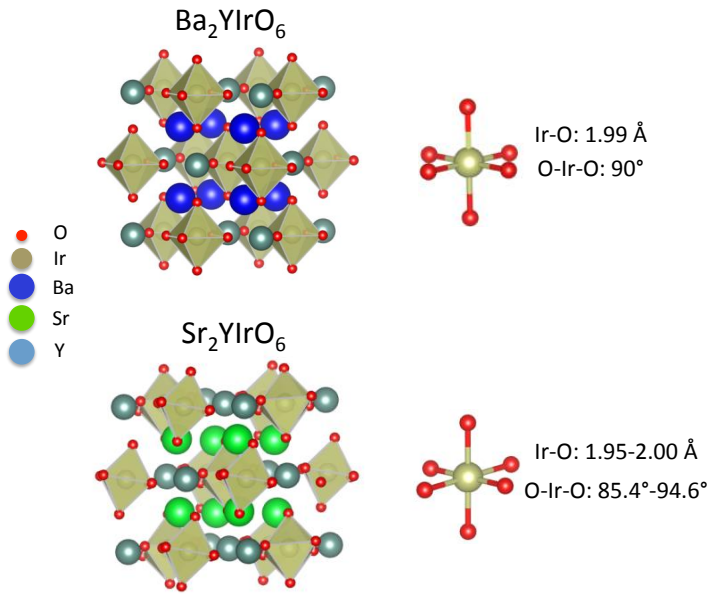

b)

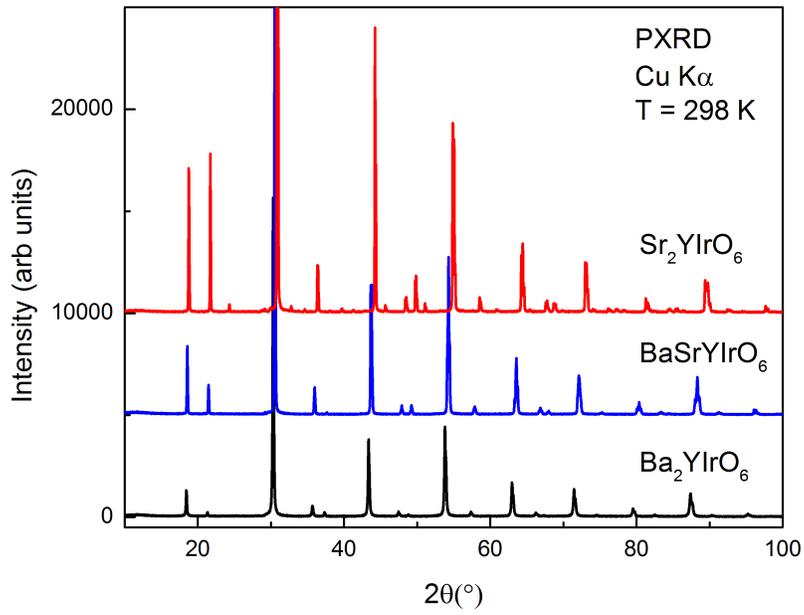

**Figure 2:**

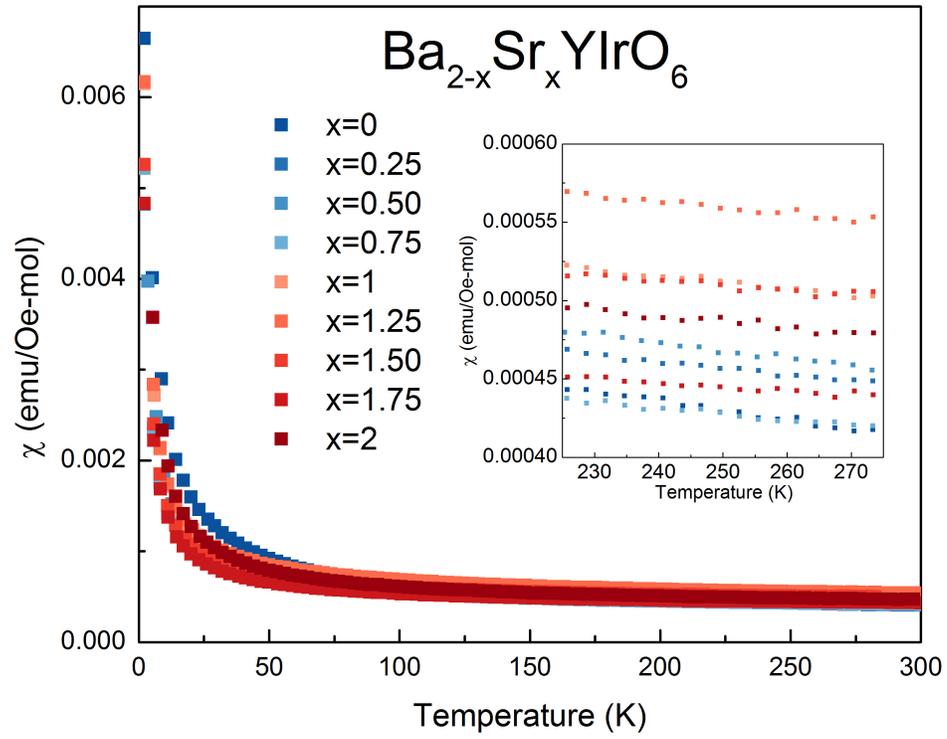

**Figure 3.**

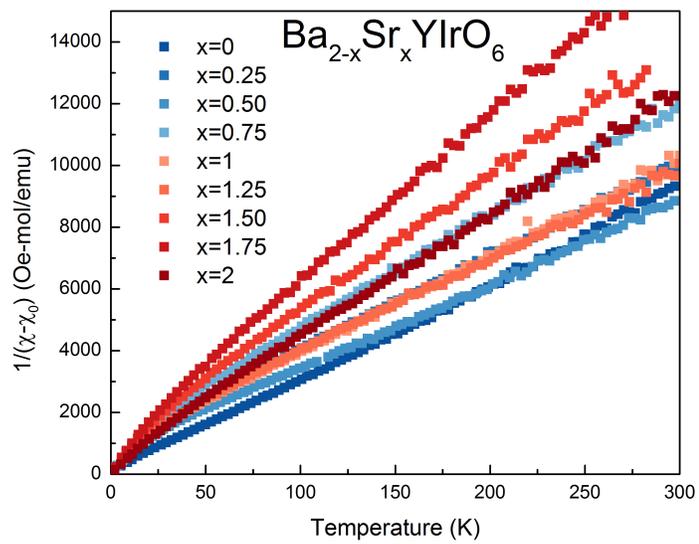

a)

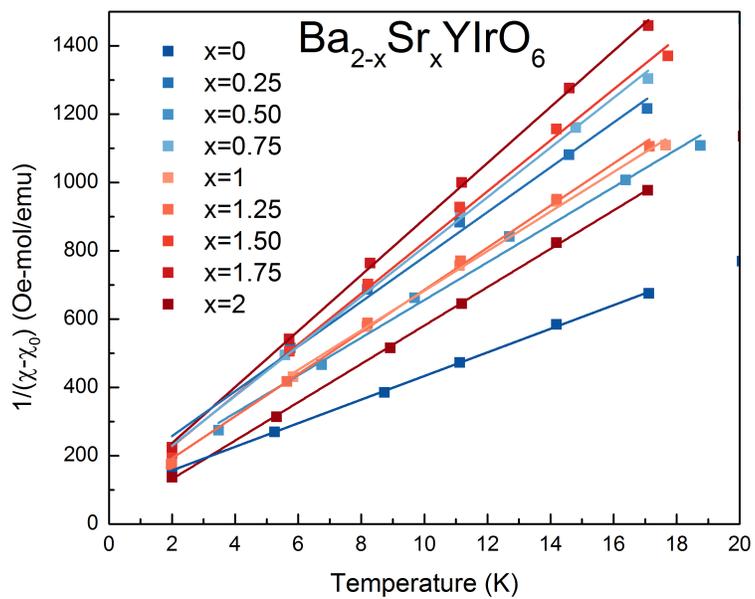

b)

**Figure 4.**

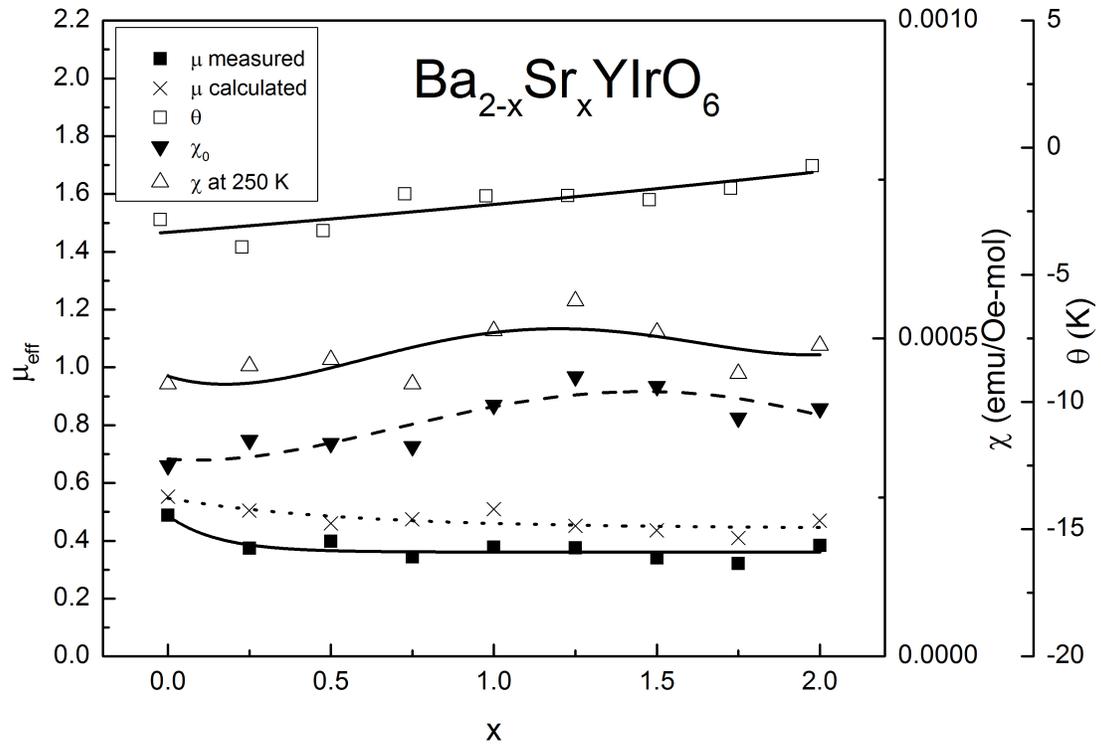

**Figure 5.**

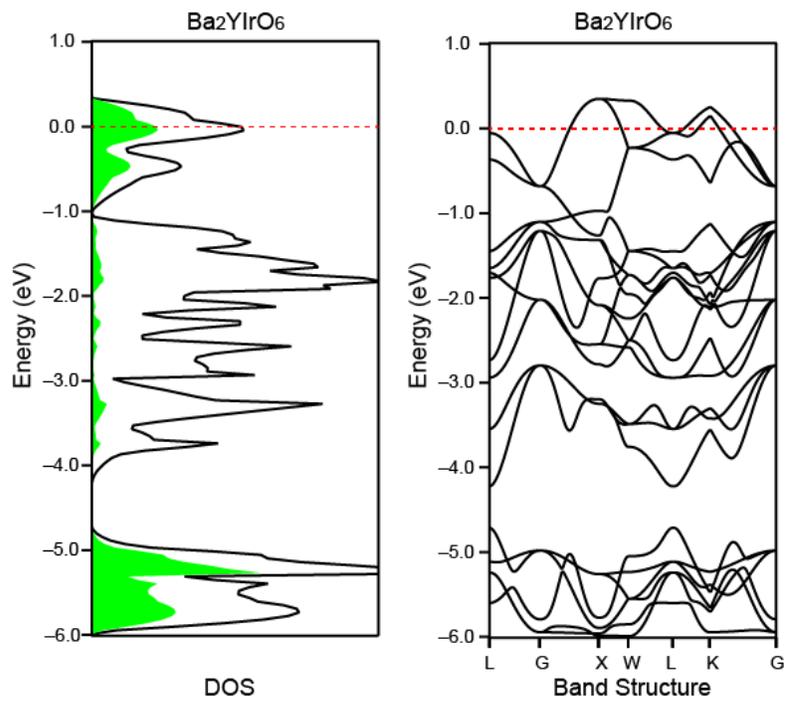

**Figure 6.**

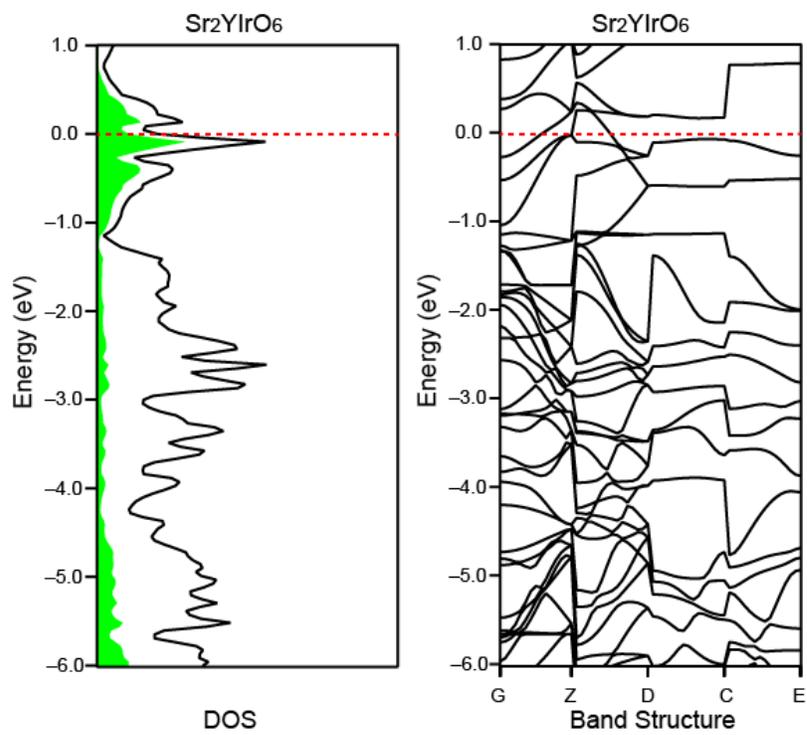